\documentclass[10pt]{iopart}
\usepackage{iopams}  
\usepackage[T1]{fontenc}
\usepackage{graphicx}	
\usepackage{amssymb}
\usepackage{xcolor}
\usepackage{bbold}

\expandafter\let\csname equation*\endcsname\relax
\expandafter\let\csname endequation*\endcsname\relax

\usepackage{amsmath}
\usepackage{hyperref}
\usepackage{amsfonts}

\usepackage{bm}
\usepackage{lmodern}

\def\x {\bm{x}}

\def\d {\rm{d}}

\def\x {\bm{x}}

\newcommand{\p}{_{\parallel}}% ||

\newcommand{\n}{{\bf \hat{n}}}
\newcommand{\HH}{\mathcal{H}\,}

\newcommand{\<}{\langle}
\renewcommand{\>}{\rangle}
\newcommand{\red}[1]{{{#1}}}
\newcommand{\blue}[1]{{{#1}}}

\newcommand\apj{The Astrophysical Journal}

\newcommand{\averageA}[1]{\left\langle #1 \right\rangle_{\Omega}}
\def\jnlref#1{{\rm#1}}
\newcommand\apjl{The Astrophysical Journal Letters}

\def\aapr{\jnlref{A\&A~Rev.}}
\def\pasp{\jnlref{PASP}}

\def\nat{\jnlref{Nature}}

\begin{document}

\title{Emergence of smooth distance and apparent magnitude in a lumpy Universe}

\author{Obinna Umeh}

\address{Institute of Cosmology \& Gravitation, University of Portsmouth, Portsmouth PO1 3FX, United Kingdom\\
Department of Physics, University of the Western Cape,
Cape Town 7535, South Africa\\
Department of Mathematics and Applied Mathematics, University of Cape Town, Rondebosch 7701, South Africa}
\ead{obinna.umeh@port.ac.uk}
\vspace{10pt}
\begin{indented}
\item[]\today
%\item[]August 2017
\end{indented}

\begin{abstract} 
The standard  interpretation of observations such as the peak apparent magnitude of Type Ia supernova made from one location in a lumpy Universe is based on the idealised Friedmann-Lemaître Robertson-Walker spacetime.  All possible corrections to this model due to inhomogeneities are usually neglected. Here, we use the result from the recent concise derivation of the area distance in an inhomogeneous universe to study the monopole and Hubble residual of the apparent magnitude of  Type Ia supernovae.  We find that at low redshifts, the background FLRW spacetime model of the apparent magnitude receives corrections due to relative velocity perturbation in the observed redshift.   We show how this velocity perturbation could contribute to a variance in the Hubble residual and how it could impact the calibration of the absolute magnitude of the Type Ia supernova in the Hubble flow. We also show that it could resolve the tension in
the determination of the Hubble rate from the baryon acoustic oscillation and local measurements. 
\end{abstract}

\vspace{1pc}

%\maketitle
%~\cite{Zhao:2021ahg}

\section{Introduction}

The Universe we observe today from a single location on Earth is not homogeneous and isotropic. This is obvious from the map of the distribution of structures in the nearby Universe by the Sloan Digital Sky Survey (SDSS)~\cite{Shi:2017qpr}. The observed distribution is clearly inhomogeneous~\cite{BOSS:2016wmc,Libeskind:2017tun}. On very large scales, that is more than 200 Mpc away from the observer, the features of isotropy in the distribution of structures begin to emerge~\cite{Aghanim:2018eyx}. Geometric homogeneity on the \blue{one} hand cannot be established independent of the Copernican principle assumption~\cite{Clarkson:2010uz,Maartens:2011yx,Clarkson:2012bg}.
Yet, when we interpret distances to or the apparent magnitudes of sources for cosmological inference, we model the underlying null geodesics using the Friedmann-Lemaître Robertson-Walker (FLRW) metric without considering the fact that null geodesics that terminate at the observer could be impacted by structures in the neighbourhood of the observer~\cite{Mustapha:1998jb}. 

This practice was motivated by the 1976 paper by Steven Weinberg~\cite{1976ApJ...208L...1W}. Weinberg showed that if conservation of photon number holds, then it is consistent to use the background FLRW spacetime on all scales and at all times. 
Weinberg's argument/proof relies on several assumptions which include: (1) the cosmological principle holds(i.e., that the average number density of photons is described in the FLRW spacetime)~\cite{Ellis:1998ha}, (2) the impact of gravity on rulers and clocks is not important~\cite{Kaiser:2015iia,Breton:2020puw}, (3) the impact of strong redial lensing effects or caustics on the luminosity distance is negligible~\cite{Ellis:1998qga,Ellis:2018led}. Also, the proof focused on the area of the screen space and not on the area/luminosity distance. Recent works have shown that the weak gravitational lensing contributes to the area/luminosity distance measurement at high redshift~\cite{Umeh:2012pn,Umeh:2014ana,Clarkson:2014pda}.

 The impact of caustics on the area distance was first studied in \cite{Ellis:1998qga} within the Swiss-cheese model of the Universe. The authors showed that the presence of radial lensing caustics leads to an increase in the area distance on average when compared to the FLRW prediction. \blue{Similarly},  it is well-known that rulers and clocks behave differently in the neighbourhood of caustics and these \blue{effects} were neglected in Weinberg's photon number conservation argument~\cite{Ellis:2018led}. Our focus here is to study the impact of radial lensing contribution to the area distance within the standard cosmology(i.e., perturbed FLRW spacetime). We work in the Conformal Newtonian gauge which is the most viable gauge for studying the dynamics of structures in the immediate neighbourhood of the observer~\cite{Clifton:2020oqx}. The details of the derivation of the area distance used in this work were presented in \cite{Umeh:2022hab}. Our target here is simply to focus on the impact of inhomogeneities on the calibration of the apparent magnitude of the Type Ia supernova (\blue{Sn~Ia}). We study how the inhomogeneities could contribute to the Hubble residuals which are usually attributed to the host galaxy properties~\cite{Kelly:2010ApJ...715..743K,SNLS:2010kps,DES:2022tgg,Dixon:2022ryo}. 
%We on the area distance and calibration of the apparent magnitude of the \blue{Sn~Ia} in the Hubble flow. 
% but the key ones are the neglect of the effect of strong gravitational lensing and 
% is a monotonic function on all scales. This is known not to apply for gravitationally bound systems such as our local group~\cite{Dunner:2006rf,Dunner:2006rc}. and finally, it assumes that the cosmological redshift can be extrapolated to zero~\cite{Umeh:2022hab}. 

 \blue{Essentially, this paper shows} that the calibration of the absolute magnitude of the \blue{Sn~Ia} in the Hubble flow with a set of local distance anchors is impacted by inhomogeneities along the line of sight. 
\blue{Furthermore, it shows} that the contribution of the inhomogeneities to the apparent magnitude of the \blue{Sn~Ia} could resolve the supernova absolute magnitude tension~\cite{Efstathiou:2021ocp}. The supernova absolute magnitude tension is another way of expressing the Hubble tension, i.e. the tension between the determinations of the Hubble rate by the early and late Universe experiments~\cite{Riess:2021jrx}. The correction to the apparent magnitude is due to the effect of radial lensing or Doppler lensing~\cite{Bacon:2014uja}. \blue{How the Doppler lensing effects could impact or contribute to the Hubble residual is also discussed}.  \blue{Finally, the estimate of the Alcock-Paczyński parameters~\cite{Alcock:1979mp} in a perturbed FLRW spacetime is provided.} It is compared to the background FLRW spacetime prediction using the cosmological fitting prescription proposed by Ellis and Stoeger in 1987~\cite{EllSto87,Umeh:2022prn}.
%We discuss how this could manifest as supernova absolute magnitude tension.~\cite{DiValentino:2021izs}.

This paper is organised as follows: In section~\ref{sec:Areadistance},  the impact inhomogeneities along the line of sight to the area distance at low redshift is provided. Section \ref{sec:impactonH0} contains discussions on how inhomogeneities impact the apparent magnitude of \blue{Sn~Ia}. For instance, in sub-section~\ref{sec:calibration}, details on how it impacts the calibration of the absolute magnitude of  \blue{Sn~Ia} is provided. How the impact of small scale inhomogeneities could provide a possible resolution of the supernova absolute magnitude tension is presented in sub-section \ref{sec:abstension}.  \blue{The impact of the small scale inhomogeneities  on the inferred Hubble rate} from the Alcock-Paczyński parameters is discussed in section \ref{sec:alcock-pay}. \blue{The argument on existence of a fundamental limit of the expanding global coordinate system is provided in section \ref{sec:breakdown} and conclusion is provided in section \ref{sec:conc}.}
For all the numerical analyses, the cosmological parameters determined from the analysis of Cosmic Microwave Background (CMB) anisotropies by Planck collaboration were used~\cite{Aghanim:2018eyx}.

\section{Impact of inhomogeneities on the area distance in cosmology}\label{sec:Areadistance}

We describe the bundle of light rays coming from a source as null geodesics and define a deviation vector as a difference between a central ray, $ \bar{x}^i$, in a bundle and the nearby null geodesic, $x^i$, at the same affine parameter, $\lambda$, as $\xi^i(\lambda) =x^i(\lambda) -   \bar{x}^i(\lambda)$.  With respect to the observer's line of sight direction, ${\n}$, $\xi^i$ maybe decomposed into two components: $\xi^i = \xi_{\p} n^i + \xi_{\perp}^i$, where $\xi_{\p} =\xi_i\, n^i$ and $\red{\xi_{\perp  }^{i} }$ are  components parallel and orthogonal to ${\n}$ respectively. 
%$N_{ij}$ is the metric on the screen space. 
The distortions experienced by the nearby geodesics due to the impart of curvature modify the area distance on an FLRW spacetime according to~\cite{Umeh:2022hab}
%\small
  \begin{eqnarray}\label{eq:Areadistance2}
  \red{{d}_A(z,{\n})}&\simeq \bar{d}_{A}(z)\bigg[1+ \frac{ \xi{\p}}{\bar{d}_{A}} + \kappa  
 -\frac{1}{4}\bigg( \gamma_{ij} \gamma^{ij}-\omega_{ij}\omega^{ij} \bigg) \bigg]
 \,,
 \end{eqnarray}
 where $\bar{d}_{A}$ is the area distance on the background FLRW spacetime,  $\kappa =- \nabla_{\perp i } \xi_{\perp  }^{i}/2 $ describes the isotropic magnification of the source image due to the weak gravitational lensing, $\gamma_{\<ij\>}= \nabla_{\perp \< i} \xi_{\perp j\>}$ is the  shear distortion, it describes the stretching of the image tangentially around the foreground mass and $\omega=\varepsilon^{ij}\nabla_{\perp [i}\xi_{ \perp j]}$ is the twist.  ${ \xi{\p}}/{\bar{d}_{A}}$  is the radial convergence~\cite{Ellis:1998qga,Bolejko:2012uj}. It is central to our work. The impact of the heliocentric peculiar motion will be discussed elsewhere, in the meantime see \cite{Kaiser:2014jca} for the estimate of the expected impact of heliocentric peculiar motion on the area distance.
 
 We work in Conformal Newtonian gauge on  a flat FLRW background spacetime:
\begin{eqnarray}\label{eq:metric}
{\d} s^2 &= a^2\big[-\big(1 + 2\Phi \big){\d} \eta^2 
 + \big(\big(1-2 \Phi\big)\delta_{ij}\red{ {\d }x^i {\d} x^j}\big]\,.
\end{eqnarray}
Here $\delta_{ij}$ is the flat metric on Minkowski spacetime, $a$ is the scale factor of the Universe and $\Phi$ is the Newtonian potential. The all-sky average  of equation \eqref{eq:Areadistance2} on the surface of constant redshift is given by%~\cite{Umeh:2022hab}
\begin{eqnarray}
\averageA{  {d}_A(z_s)}&=\bar{d}_A\bigg[1+  \averageA{\frac{\xi{\p}}{\bar{d}_A} }
-\frac{1}{2}\averageA{ \gamma_{ij} {\gamma}^{ij}}\bigg]\,.
\label{eq:averageDA}
\end{eqnarray}
We made use of Gauss's theorem to set the all-sky average of $\kappa$ to zero $\averageA{\kappa} = 0$.  $\omega_{ij}\omega^{ij}$ vanishes at second order in perturbation theory and $\averageA{ \gamma_{ij} {\gamma}^{ij}}$ becomes important only at high redshift and its total contribution is about a few percent at $z\ge1$~\cite{Clarkson:2014pda}. Hence, we do not consider the weak gravitational lensing correction in the discuss that follows. 
 
The leading order correction to the area distance at  low redshift is given by the second term in equation \eqref{eq:averageDA}, i.e. $ \averageA{{\xi{\p}}/{\bar{d}_A} }$. This correction comes from  the perturbation of the  redshift,  $\delta z$, due to Doppler effect~\cite{Bacon:2014uja,Umeh:2022hab}. When interpreting distances to sources on constant redshift surface(i.e., area distance as a function of the cosmological redshift), the monopole of the radial lensing correction, $ \averageA{{\xi{\p}}/{\bar{d}_A} }$, translates to an increase in distance to the source, $\delta r = \delta z/\HH$, the dominant term  in $\delta z$ is the relative peculiar velocity; 
\begin{eqnarray}
\red{\delta z(z,{\n}) \approx \partial_{\|}{v_s}-  \partial_{\|}v_o +  \mathcal{O} \left(\Phi_o -\Phi_s\right) \,,}
\end{eqnarray}
 where $\partial_{\|}v_o$ and $\partial_{\|}v_s$ are the peculiar velocities of the observer and the source respectively,  $\HH$ is the conformal Hubble rate. At the linear order in perturbation theory, this term may be evaluated on the background spacetime, this is the well-known Born approximation~\cite{1926ZPhy...38..803B}, however, at the second order, we need to include the post-Born correction terms~\cite{Nielsen:2016ldx}: 
\begin{eqnarray}
\delta z({\bar{r}}, {\bar{\theta}^i}) &\to& \delta z({r}_s-\delta r_s, {\theta}^i-\delta\theta^i_s)\,,
\\
& =&\delta z({{r}_s}, {{\theta}^i_s})  - \delta r_s\partial_{\p}\delta z({\bar{r}}, {\bar{\theta}^i}) - \delta\theta^i_s \nabla_{\perp}\delta z({\bar{r}}, {\bar{\theta}^i}) \,, 
\end{eqnarray}
where,  $\delta\theta^i_s $ is the angular perturbation. $\delta\theta^i_s \nabla_{\perp}\delta z$ captures the effect of gravitational deflection, it is not important to our discussion because it is sub-dominant at low redshift. Our focus is on the radial correction, $ \delta r_s\partial_{\p}\delta z $, it dominates at very small redshifts~\cite{Umeh:2022hab}
\begin{eqnarray}\label{eq:everyleadingterm}
  \averageA{\frac{\xi{\p}(z_s)}{\bar{d}_A} } &\approx&  \frac{1}{r_s\HH^2_{s} }  \bigg[
\averageA{\partial_{||} v_{s} \partial^2_{||} v_{s}}  - \averageA{\partial_{\|}{v_o} \partial^2\p{v_s}}\bigg]\,,
\end{eqnarray}
where $r_s$ is the comoving distance to the source. The all-sky average of $\averageA{\partial_{||} v_{s} \partial^2_{||} v_{s}} $ vanishes, this implies that the key  leading order contribution to  $\averageA{{\xi{\p}}/{\bar{d}_A} }$ is given by~\cite{Umeh:2022hab}
\begin{eqnarray}\label{eq:lowz_correction}
  \averageA{\frac{\xi{\p}(z_s)}{\bar{d}_A} }&\approx % - \frac{1}{\chi_s \HH_s}\delta z\approx 
 -  \frac{1}{r_s\HH^2_{s} } \averageA{   \partial_{\|}v_o \partial^2\p{v_s}} \,.
\end{eqnarray}
Equation \eqref{eq:lowz_correction} is the leading order term in the expression for the area distance given in equation \eqref{eq:averageDA}. The coupling between the peculiar velocity term evaluated at the observer position and the Kaiser redshift space distortion term evaluated at the observer suggests that this term describes the "parallax effect". That is the apparent displacement in the background source position as the observer move relative to the local over-density.  Astronomers utilise this effect as a tool to calculate distances to both galactic and extragalactic  sources~\cite{Riess:2018byc,Paine:2019vep}.

It is straightforward to expand equation \eqref{eq:lowz_correction} in spherical harmonics and estimate the ensemble average if the peculiar velocity term (i.e., $\partial_{\parallel}v_o$) is known. In our case, it corresponds to the peculiar velocity of the heliocentric observer. Our perturbation theory formalism is not valid on the solar system's scale~\cite{Macaulay:2011av}. Therefore, to handle this correctly, we borrow insights from the analysis of the temperature anisotropies of the Cosmic Microwave Background radiation and expand equation \eqref{eq:everyleadingterm} in full-sky spherical harmonics and drop the first two harmonics $\ell =0,$ $\ell =1$ which are coordinate dependent(see~\cite{Hu:1997hp,Umeh:2022hab} for detail on how to do this).  The impact of $\ell =0$, $\ell =1$ harmonics are then included by Lorentz boosting to the Heliocentric frame of reference~\cite{Kaiser:2015ada,Planck:2020qil}.  The monopole of coordinate independent part of equation \eqref{eq:everyleadingterm} is given by~\cite{Umeh:2022hab}
%The full sky spherical harmonics decomposition of equation  \eqref{eq:lowz_correction} is givenby~\cite{Umeh:2022hab}
\begin{align}\label{eq:harmonic-decomp}
 \averageA{ \frac{\xi{\p}(z_s)}{\bar{d}_{A}} }&\approx&
\frac{1}{r_s}( f(z_s)D(z_s))^2 
%\\ \nonumber &\times 
\sum_{\ell =2}^{\ell_{\rm{max}}} (2\ell+1) \int \frac{d k_1}{2\pi^2} k_{1}P_{m}(k_1)
 j'_{\ell}(k_1  r_s)j''_{\ell}(k_1  r_s)\,,
 %-  \frac{  \HH_o}{r_s \HH_s}(Df)(D_o f_o)\qquad
 % \\ \nonumber &
%\times   \sum_{\ell = 0}^{\ell_{\rm{max}}} (2\ell+1)\int \frac{{\d} k}{2\pi^2} k P_{m}(k)  j'_{\ell}(kr_o) j''_{\ell}(kr)\,,
\end{align}
where $j_{\ell}$ is the spherical Bessel function of order $\ell$, \blue{$'$ on $j_{\ell}(k r)$ denotes derivative wrt the argument, i.e. $(kr)$}, $f$ is the rate of growth of structure,  $P_{m}$ is the matter density power spectrum, $k$ is amplitude of the wave vector $k = |{\bm{k}}|$.
For the results we discuss here, we consider the linear perturbation theory prediction of $P_{m}$ and the halo model prediction~\cite{Cooray:2002dia,Mead:2015yca}.  When we quote percentage correction, we refer to only the results obtained using the linear matter perturbation theory prediction for $P_{m}$. 

  \begin{figure}[h]
\centering 
\includegraphics[width=140mm,height=100mm] {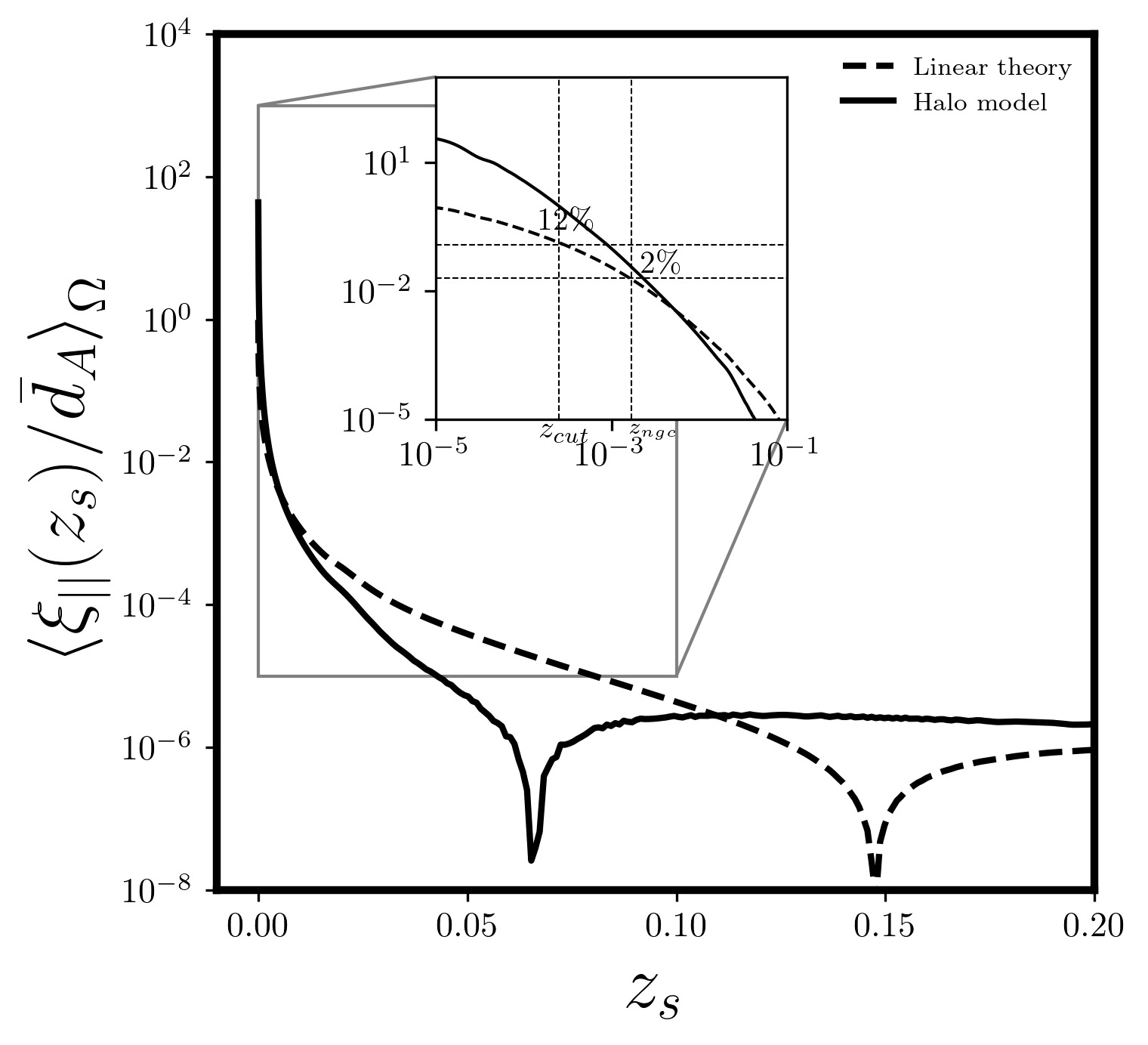}
\caption{\label{fig:BiasEv}\red{We show the monopole of the radial lensing contribution $\averageA{ {\xi{\p}}/{\bar{d}_{A}} }$ to  the area distance. At low redshifts, this is proportional  to the fractional difference between the monopole of the area distance and the background FLRW  expectation: $\averageA{ {\xi{\p}}/{\bar{d}_{A}} } \approx (\averageA{  {d}_A}- \bar{d}_{A})/\bar{d}_{A}$. The legend shows the dependence of the size of the correction on the model {used to estimate} the matter power spectrum(linear perturbation theory (dashed line) and halo model (thick line)).  For the inset, we zoomed in on the very small redshifts $(z < 0.01)$. 
We identified two redshift positions on the inset: $z_{\rm{cut}}$ and $z_{\rm{ngc}}$. 
The $z_{\rm{cut}}$ corresponds to the minimum redshift below which the spacetime is not expanding (we describe    how to determine $z_{\rm{cut}}$ in detail in section \ref{sec:breakdown}.).  
From the analysis, we find that $z_{\rm{cut}}= 2.4\times 10^{-4}$. At $z_{\rm{cut}}$, we find about 12\% correction to the area distance based on the background FLRW spacetime. Furthermore, $z_{\rm{ngc}}$ is the cosmological redshift for the NGC 4258. We find about 2\% correction to the area distance at $z_{\rm{ngc}} = 0.0018$. NGC 4258 is one of the key local distance anchors used to calibrate the co-located cepheids(cepheids in the same host as the \blue{Sn~Ia}). The co-located cepheids are then used to calibrate the \blue{Sn~Ia} in the Hubble flow.   In the numerical computation, we find that it is enough to set $\ell_{\rm{max}} $ to $\ell_{\rm{max}} =100$.}
  }
\end{figure}

\blue{The correction to the area distance is shown} in figure \ref{fig:BiasEv}. It is clear that at $z \ge 0.023$, the corrections to the background FLRW model are less than $0.1\%$, hence the background FLRW spacetime model is valid. However, for $z \le 0.01$, the correction to the background could exceed 12\% at $z_{\rm{cut}}$. Here, the $z_{\rm{cut}}$ is the minimum value the cosmological redshift can take in an expanding universe.  It corresponds to the minimum length scale below which the spacetime is not expanding.  The $z_{\rm{cut}}\le z \le 0.023$ redshift range corresponds to the first and second rungs of the cosmic distance ladder according to the SH0ES collaboration classification scheme~\cite{Riess:2016jrr}. As we shall see in the subsequent section, the apparent optical properties of sources in $z_{\rm{cut}}\le z \le 0.023$ is very important, they play a crucial role in the calibration of the absolute magnitude of the \blue{Sn~Ia} in the Hubble flow. 
%Among all the distance anchors used by the SH0ES collaboration, we can only predict the average distance to NGC 4258 using the cosmological perturbation theory technique~\cite{Riess:2021jrx}. Irrespective of this constraint, the perturbation theory tools are valid to calculate the apparent magnitude of the \blue{Sn~Ia} down to $z = z_{\rm{cut}}$.
%Note that the momentum integral in equation \eqref{eq:harmonic-decomp} is cuff-off independent.
%In summary,  we find that the smooth area distance is given by

\blue{The most essential point from the section is that }the effective area distance model in the low redshift limit in the presence of inhomogeneities is given by
\begin{eqnarray}\label{eq:averageareadistance}
\averageA{  {d}_A (z_s)}
&= \frac{{d}^{\rm{eff}}_{H} (z_s)}{(1+z_s)} \int _{z_{\rm{cut}}}^{z_s}\frac{dz'}{{\sqrt {\Omega _{m}(1+z')^{3}+1-\Omega_{m0}}}}\,,% \rightarrow \lim_{z\to0} \averageA{  {d}_A(z,{\n})} 
\end{eqnarray}
where  $\Omega_{m0}$ is the matter density parameter,   ${d}^{\rm{eff}}_{H}= d_{H} \left[1 +  \averageA{ {\xi{\p}}/{\bar{d}_{A}} } \right]$ %$\approx d_{H}\left[1+{2}\averageA{{ \sigma_{ij}\sigma^{ij}}/{{15}H^2_0}} \right]$
 is the effective Hubble distance and  $d_{H}=c/{H}_0$ is the global Hubble distance.  Using this model for the distance to NGC 4258 \blue{gives} about 2\% correction to the prediction of the background FLRW model. 
 In the next section, \blue{this point is investigated in greater detail in order to understand }how it impacts the apparent  and absolute magnitude of the \blue{Sn~Ia} in the Hubble flow. 
 %, $d_L$ via the distance duality relation $d_L =(1+z)^2d_A$ impacts the calibration of the SN1A in the Hubble flow.

\section{Cosmology in the presence of inhomogeneities }\label{sec:impactonH0}

\subsection{Apparent magnitude and calibration of the cosmic distance ladder}\label{sec:calibration}

The apparent magnitude, $m_b$, of the \blue{Sn~Ia} is a function of the bandpass (filter), the spectral flux density and the fundamental standard spectral flux density. Our interest is on the effect of the small scale inhomogeneities on the luminosity distance, $d_L$.  \red{Assuming photon number conservation, we can obtain $d_L$ from $d_A$ which we have already calculated using the distance duality relation: $d_L =(1+z)^2d_A$.}
% we focus on the bolometric apparent magnitude of the \blue{Sn~Ia}, $m_b$. It is straightforward to include the bandpass corrections~\cite{Hogg2022arXiv}. 
Within a given filter, $m_b$ is related to the flux density $F_{d_L}$ according to %at a given luminosity distance $d_{L}$ in a given spectral filter 
\begin{eqnarray}\label{eq:apparentmag}
m_b=-2.5\log _{10}\left[F_{d_{L}}\right]\,.
\end{eqnarray}
For cosmological inference, the apparent magnitude is useless without calibration. The peak apparent magnitude of the \blue{Sn~Ia} in the Hubble flow is calibrated using secondaries, for example, the Tip of the Red Giant Branch (TRGB)~\cite{Freedman:2000cf}, cepheids~\cite{Reid:2019tiq}, etc. The essence of calibration is to determine 
the absolute magnitude ($M_b)$ of the \blue{Sn~Ia}. The absolute magnitude is defined as the apparent magnitude measured at a distance $D_{F}$: $M_b=-2.5\log _{10}\left[{F_{D_F}}\right],$
where $F_{D_F}$ is called the reference flux or the zero-point of the filter~\cite{Rizzi:2007ni,2015PASP..127..102M,Madore:2021ktx}. The observed flux density at both distances obeys the inverse square law with the luminosity distance:
${F_{d_L}}/{ F_{D_{F}}} = \left[{D_{F}}/{d_L}\right]^2$. The cosmological magnitude-redshift relation or the distance modulus is defined by setting $D_{F} =1 $[Mpc] 
\begin{eqnarray}\label{eq:mag-redshift-relation}
\mu= m_b - M_b =  5 \log_{10} \left[\frac{d_{L}}{[\rm{Mpc}]}\right] +25\,,% = 5 \log \left[\frac{d_{L}}{[\rm{pc}]}\right] - 5
\end{eqnarray}
where  $\mu$ is the distance modulus.  \red{The SH0ES collaboration calibrates the absolute magnitude of the \blue{Sn~Ia} in the Hubble flow using a sub-sample of nearby \blue{Sn~Ia} co-located with cepheids in the second rung of distance ladder:}
\begin{eqnarray}\
\red{M_b = \blue{m_{0,\rm{SnIa}}} - \mu_{0,\rm{ceph}}\,,}
\end{eqnarray}
where \blue{$m_{0,\rm{SnIa}}$ } is corrected-peak apparent luminosity of \blue{Sn~Ia} in the same host as cepheid, $ \mu_{0,\rm{ceph}}$ is the distance modulus of the co-located cepheid.    $ \mu_{0,\rm{ceph}}$ is obtained from the calibrated Leavitt law~\cite{1930AnHar..85..143L}
\begin{eqnarray}
m_{0,{\rm{ceph}}} = \mu_{0,{\rm{ceph}}} + M_{H,{\rm{ceph}}} + b_w\left( \red{\log_{10} }P -1) +  Z_{W} \left[O/H\right]\right]\,,
\end{eqnarray}
where $m_{o,{\rm{ceph}}} $ is the apparent magnitude of Cepheids, $ \left[O/H\right]$ is the metallicity correction, $P$ is the period of the host measured in days, $M_{H,\rm{ceph}}$ is the fiducial absolute magnitude of cepheid evaluated when \red{$\log_{10} P = 1$} or $P =10$ days and the metallicity correction set to the solar metallicity. $b_W$ and $Z_W$ define the empirical relation between the cepheid period, metallicity and luminosity.

The recent SH0ES analysis made use of an independent geometric distance estimate to NGC 4258 as an anchor to calibrate $M_{H,\rm{ceph}}$ for cepheids in NGC 4258~\cite{Riess:2021jrx} 
\begin{eqnarray}
 \red{M_{H,\rm{ceph} } = m_{\rm{NGC}} - \mu_{\rm{NGC}}\,,}
\end{eqnarray}
where, $m_{\rm{NGC}} $ is the corrected apparent magnitude of cepheids in NGC 4258 and $\mu_{\rm{NGC}}= 5\log_{10} [d_{\rm{NGC}}/{\rm{[Mpc]}}]$ is the independent distance modulus to the NGC 4258. 
\red{The NGC 4258 is an intermediate mass spiral galaxy with  a central supermassive black hole. It is about 7.5 [Mpc] away from earth with a cosmological redshift of about $z\approx 0.0018$.}
The distance estimate to NGC 4258 (i.e $d_{\rm{NGC}}$) is obtained from the shape of the MASER luminosity profile in the accretion disk of the supermassive black hole at the centre of NGC 4258. From the width of the luminosity profile, $d_{\rm{NGC}}$ is given by 
\begin{eqnarray}
\red{d_{\rm{NGC}} =\frac{ \partial_{\theta}v_{\rm{LoS}}}{a_{\rm{LOS }}}\,,}
\end{eqnarray}
 where  $v_{\rm{LoS}}$ is the Line of Sight (LoS) velocity to the localised MASER emission around the supermassive black hole, $\partial_{\theta} v_{\rm{LoS}}$ is the gradient of $v_{\rm{LoS}}$ on the sky, $a_{\rm{LOS }}$ is the LoS acceleration and $\theta$ is the angle on the sky as measured by the observer~\cite{Greene:2021shv}. %It is usually set to $1 {\rm{Mpc}}$, this is fine for a homogeneous matter distribution~\cite{Lombriser:2019ahl}.   The NGC 4258 is about 7.5 Mpc away. It lies in the regime where we found about 10\% correction to the distance in figure \ref{fig:BiasEv}.  
  
  Other geometric distance anchors include the Milky-Way Cepheids~\cite{Riess:2018byc}. In this case, the parallax method is used to estimate the distance to the Milky-Way Cepheids and the Large Magellanic Cloud (LMC), where the  distance  is obtained from the dynamics of the detached eclipsing binary systems~\cite{2019Natur.567..200P}.   
Both of the distances to LMC and the milky-way cepheids are less than 1[Mpc] away.  \red{The cosmological perturbation theory technique is inadequate for estimating distances to objects that are less than 1 [Mpc] away from the observer \cite{Umeh:2022hab}. That is $d_{A} \approx d_L < 1$[Mpc] is non-perturbative.   As a result, this study focuses on NGC 4258 which is about 7.5 [Mpc] away. In general, for  $d_{A} \approx d_L \ge 1$[Mpc], the cosmological perturbation theory technique provides a valid description of the impact of inhomogeneities.}
 
 %and we cannot describe how inhomogeneities impacts these on average using the cosmological perturbation theory techniques(we shall explain this in detail in section \ref{sec:breakdown}.). 
%  \red{ The essential point is that for $d_{A} \approx d_L \ge 1$[Mpc], our cosmological perturbation theory technique provides a valid description of the impact of inhomogeneities on distances to the anchors}. 
 %In all of these, the most important criteria is the ability to predict the luminosity distance at about $d_L\ge 1 $[Mpc]. From figure \ref{fig:BiasEv}, we can do this using the cosmological perturbation theory technique.
  %There are plans to extends this to about 80 Mpc volume with JWST\cite{Dhawan:2022yws}. This will be holy grail for verifying the local dependence of the distance measurement we claim here. 
%

\subsection{Hubble residuals and inhomogeneities along the line of sight}\label{sec:Hubble residuals}

  There has been a huge interest in understanding  the observed Hubble residuals in the samples of the \blue{Sn~Ia}~\cite{DES:2022tgg,Dixon:2022ryo}. The Hubble residual is defined as the difference between the distance modulus of the corrected peak apparent magnitude of the \blue{Sn~Ia} inferred from the light curves as a function of the \blue{Sn~Ia} host cosmological redshift and the apparent magnitude calculated assuming the background FLRW spacetime predication for the luminosity distance~\cite{Kelly:2010ApJ...715..743K,Brout:2022vxf}.  Correlations have been found between the Hubble residuals and host galaxy properties such as the size, stellar mass, ${\rm{H}}\beta$ content etc~\cite{Campbell:2016zzh,Dixon:2022ryo,DES:2022tgg}. 
  %A recent study has reported a non-negligible impact of the correlation on the measurement of cosmological parameters~\cite{Brout:2022vxf}. For example, correlations in the Hubble residuals leads to negative values in the equation of state parameter $\omega$ and a shift in the matter density parameter $\Omega_m$ to lower values~\cite{Campbell:2016zzh,Kim:2019npy}.  

Using the cosmological perturbation theory technique, we can calculate the Hubble residuals provided  that the calibration of the absolute magnitude \red{is} held fixed. For example, keeping the calibration of $M$ fixed, we find that the Hubble residuals is given by
\begin{eqnarray}
\averageA{\Delta \mu(z_s)} \equiv  \averageA{\mu} - \bar{\mu} &=& 5 \log_{10} \left[ 1 + \averageA{ \frac{\xi{\p}}{\bar{d}_{A}}} - \frac{1}{2} \averageA{ \frac{\xi^2{\p}}{\bar{d}_{A}^2} }\right] \,,
\\
&=&\red{\frac{5}{ \ln({10})} \left[  \averageA{ \frac{\xi{\p}}{\bar{d}_{A}}} - \frac{1}{2} \averageA{ \frac{\xi^2{\p}}{\bar{d}_{A}^2} }\right] }\,,
\end{eqnarray}
 where the spherical harmonic expansion  of $\averageA{{\xi^2{\p}}/{\bar{d}_{A}^2}}$ is given by
\begin{eqnarray}\label{eq:rgoesq}
{ \averageA{\frac{\xi^2{\p}}{\bar{d}_{A}^2}}}  &\approx& \frac{f^2(z_s)D^2(z_s)}{r_{s}^2} 
\sum_{\ell = 2}^{\ell_{\rm{max}}} (2\ell+1)
%\\ \nonumber &&\times
\int \frac{{\d} k}{2\pi^2}  P_{m}(k)  j'_{\ell}(kr_s) j'_{\ell}(kr_s)\,.%\qquad
\end{eqnarray}
The variance(error)  in $\Delta \mu$ due to the radial lensing effect maybe estimated using 
\begin{eqnarray}
\red{\sigma^2_{\Delta \mu}=  \averageA{(\Delta \mu)^2}  -  \averageA{\Delta \mu}^2,}
\end{eqnarray}
The standard deviation is obtained from $ \red{\sigma^2_{\Delta \mu}: \sigma_{\Delta \mu} =\sqrt{\sigma^2_{\Delta \mu}}}$~\cite{Umeh:2010pr}.
The full sky harmonic expansion of $ \averageA{(\Delta \mu)^2}$ is given by
\begin{eqnarray}\label{eq:musqaverage}
\red{ \averageA{(\Delta \mu)^2} \approx \left[\frac{5}{\ln({10})} \right]^2 \left[ \averageA{ \frac{\xi^2{\p}}{\bar{d}_{A}^2} }\right]\,.}
 \end{eqnarray}
    \begin{figure}[h]
\centering 
\includegraphics[width=100mm,height=70mm] {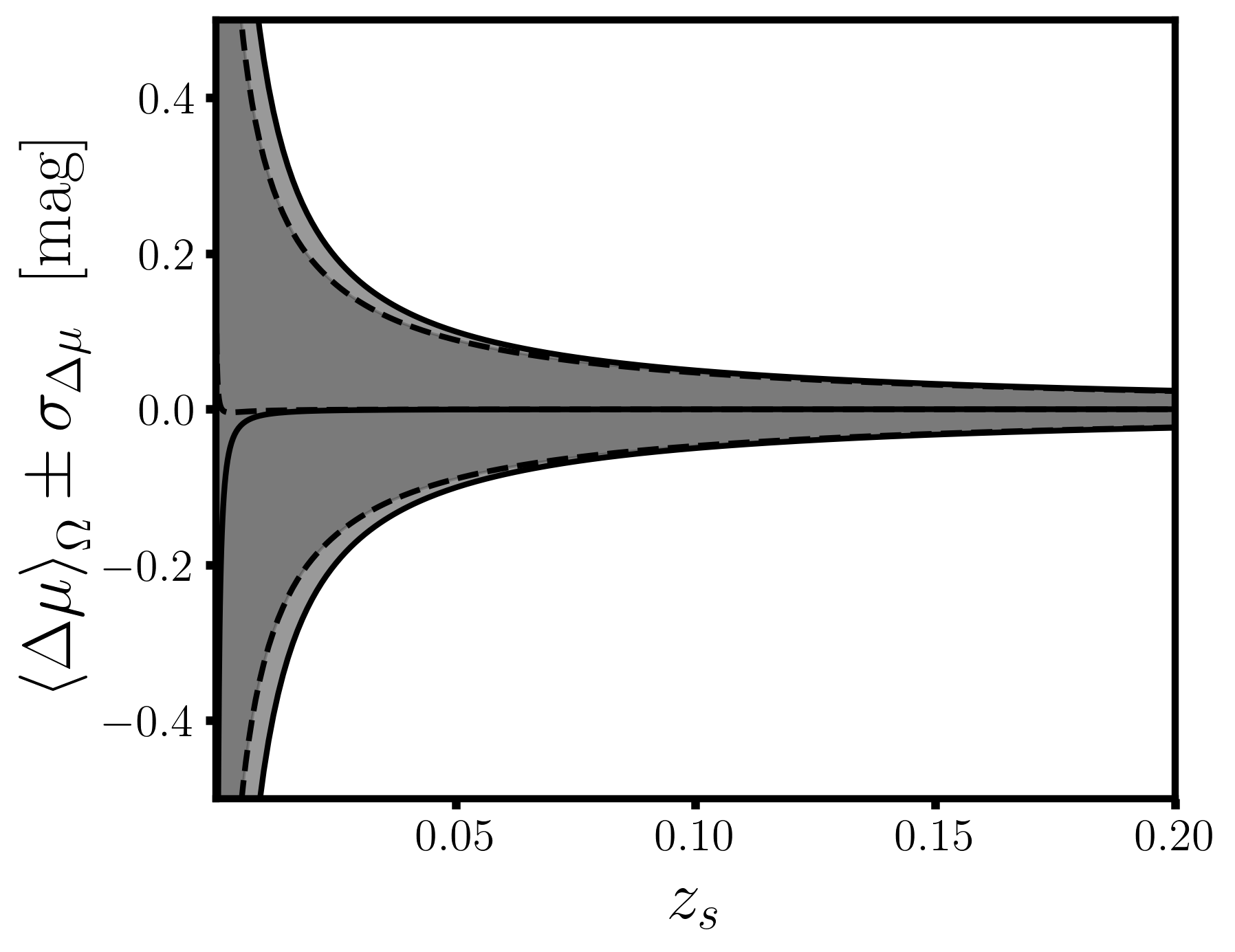}
\caption{\label{fig:Deltamu}\red{\blue{This is} the contribution to the Hubble residual due to the small scale inhomogeneities along the line of sight as a function of the cosmological redshift.  For the dashed lines, we made use of the linear power spectrum and for the thick lines, we used the halo model estimate of the matter power spectrum. 
The shaded area corresponds to the range of expected standard deviation due to small scale inhomogeneities. }
%Using the Halo power spectrum increases the spread at intermediate distances.
  }
\end{figure}
\blue{The  plot} of the Hubble residual plus or minus the expected error from the small scale inhomogenieties as a function of the cosmological redshift is shown in figure \ref{fig:Deltamu}. 
\red{As expected, the effect of the small scale  inhomogeneities contribute most significantly to the monopole of the Hubble residuals in the regions where the distance anchors live (i.e., very low redshifts).}  However, the impact of the variance extends to intermediate redshifts. Figure \ref{fig:Deltamu} also shows that the impact of effect of small scale inhomogeneities along the line of sight could account for a part or majority of the observed Hubble residuals. More detailed study would be required to account for all the expected variance in the Hubble residual especially at high redshifts. \red{Note that the effects of tangential weak gravitational lensing was neglected in this analysis, it will become important at high redshifts.}  \red{Also, the improved modelling of the peculiar velocity would be  required in order to account for the effect of the bulk flow.}
%The contribution of tangential weak gravitational lensing that we neglected is expected contribute to equation \eqref{eq:musqaverage} at high redshift.

\subsection{Resolution of the supernova absolute magnitude tension}\label{sec:abstension}

It is possible to calibrate the absolute magnitude of the \blue{Sn~Ia} in the Hubble flow using the sound horizon scale at the surface of the last scattering, $r_\star$, as an anchor~\cite{Camarena:2019rmj}. This approach is known as the inverse distance ladder approach. The approach assumes that the distance duality relation holds and that the background FLRW spacetime predication for the luminosity distance is valid on all scales~\cite{Cuesta:2014asa}. 

Using the inverse distance ladder technique and the CMB-Planck's constraint on $r_\star$, the authors of \cite{Camarena:2019rmj} found that the absolute magnitude for pantheon \blue{Sn~Ia} samples in the Hubble flow to be $ M^{\rm{P18}} = - 19.387\pm 0.021\,\, {\rm{mag}}$. However, using the distance estimate to NGC 4258 as described in sub-section \ref{sec:calibration}, the SH0ES collaboration found the absolute magnitude for the same sample to be $M^{\rm{SH0ES}} = -19.269 \pm0.029$~\cite{Riess:2021jrx}. Using different local distance anchors lead to a value which is consistent with this.  The difference between the use of early Universe sound horizon as an anchor and the late Universe local distance anchors is given by $M^{\rm{SH0ES}} - M^{\rm{P18}} = 0.118 \pm 0.035 \, [{\rm{mag}}]$. This difference is known as the supernova absolute magnitude tension~\cite{Efstathiou:2021ocp}.

Using the flux inverse square law $F_{d_L} = L/ 4\pi d_L^2$ which holds in any spacetime and keeping the intrinsic luminosity, $L$ fixed, equation \eqref{eq:apparentmag} can be written as
\begin{eqnarray}
\averageA{\Delta m(z_s)}  &\equiv & \averageA{m_b}- \bar{m}_b\,,
\\
 &=& 5 \log_{10} \left[ 1 + \averageA{ \frac{\xi{\p}}{\bar{d}_{A}}} - \frac{1}{2} \averageA{ \frac{\xi^2{\p}}{\bar{d}_{A}^2} }\right] \,,
 \label{eq:difference}
\end{eqnarray}
where  $\bar{m}_b = -2.5 \log_{10} L +  5 \log_{10}\bar{d}_{L}$ is the apparent magnitude with the luminosity  distance given by the background FLRW spacetime. This is in line with the early Universe approach.
%Similarly, we estimate the variance in $\Delta m$ using ${\rm{Var}} \left[\Delta m\right] =  \averageA{(\Delta m)^2}  -  \averageA{\Delta m}^2, where $ \averageA{(\Delta m)^2}$ is given by
%\begin{eqnarray}
% $\averageA{(\Delta m)^2} = 57.5646 \log_{10} \left(1 +  \averageA{ {\xi^2{\p}}/{\bar{d}_{A}^2} }\right)\,.$
%\end{eqnarray}
  \begin{figure}[h]
\centering 
\includegraphics[width=100mm,height=70mm] {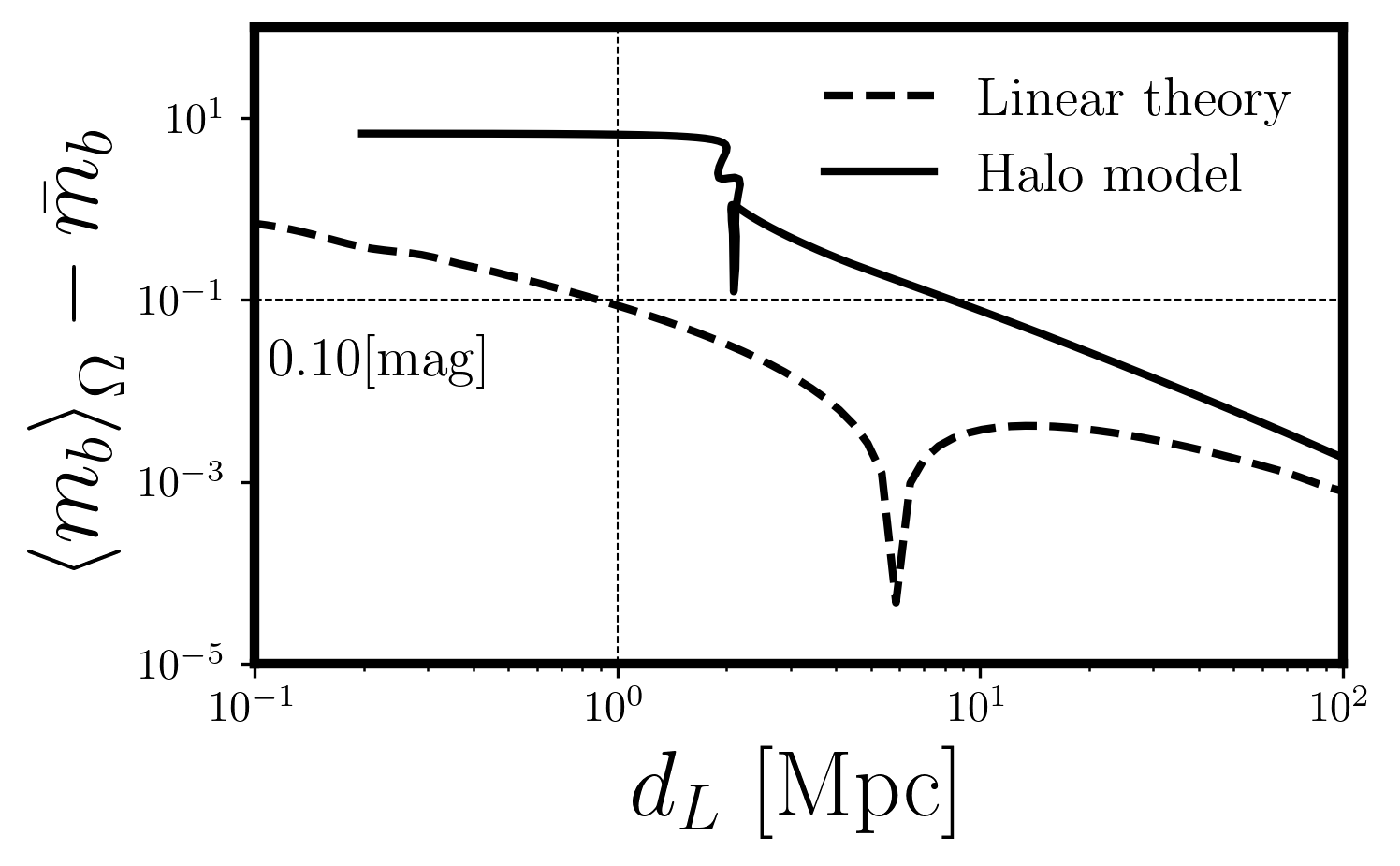}
\caption{\label{fig:differenceapparenet} The plot above shows the difference between the monopole of the apparent magnitude and the apparent magnitude predicted by the background FLRW spacetime as function of the luminosity distance.  {The dashed vertical line at $d_L = 1$[Mpc], indicates a  difference  of about $0.10$ [mag] at the corresponding horizontal intersection.} This is where the absolute magnitude of the \blue{Sne~Ia} in the Hubble flow are determined.   The position where the curve goes through zero will likely disappear when the next to leading order terms are added.  {The dashed and thick lines indicate the cosmological perturbation theory and halo model predication for the matter power spectrum respectively. }
%the local and early Universe determined absolute magnitude of the \blue{Sn~Ia} (i,e $M^{\rm{SH0ES}}  - M^{\rm{P18}}  = 0.118 \pm 0.035 \, [{\rm{mag}}]$). 
  }
\end{figure}
Now, recall that the local determination of the supernova absolute magnitude does not assume a model for the luminosity distance when this calibration is done. Within our perturbation theory framework, we can calculate the monopole of the apparent magnitude down to $1$ [Mpc]. Therefore, our claim is that our calculation of the monopole of the apparent magnitude i.e. \red{ $ \averageA{m_b}$} in a perturbed FLRW spacetime gives a better representation of what is measured than the background FLRW model on scales where the absolute magnitude is determined.   
Evaluating the difference in apparent magnitudes at 1 [Mpc] gives the measured difference in supernova absolute magnitude between the late and early Universe approaches. 

It is clear from figure \ref{fig:differenceapparenet} that by including the effect of inhomogeneities on the apparent magnitude, the difference can easily be accounted for. In other words, if the early Universe calibration of the supernova absolute magnitude uses equation \eqref{eq:averageareadistance} instead of the background FLRW spacetime equivalent, the supernova absolute magnitude tension could be resolved. 
We can see this by calculating this difference explicitly using the monopole of the generalised distance modulus 
\begin{eqnarray}
\averageA{m_b}-\averageA{M}^{R} &=& 5 \log_{10} \bar{d}_{L} + 25 \,.
\end{eqnarray}
We made use of the general expression for the luminosity distance $d_{L} = \bar{d}_{L} \left[1+ \xi_{\p}/\bar{d}_A\right]$ and have absorbed the correction due to inhomogeneities  into what we  call the renormalised absolute magnitude(this corresponds to the supernova absolute magnitude measured by the SH0ES collaboration)
\begin{eqnarray}\label{eq:renormM}
\averageA{M}^{R}-M_b &=& \frac{5}{\ln({10})} \left[\averageA{ \frac{\xi{\p}}{\bar{d}_{A}}} - \frac{1}{2} \averageA{ \frac{\xi_{\p}^2}{\bar{d}_{A}^2} }\right]_{d_{L}  = 1 [\rm{Mpc}]} \,,
\\
&\approx & \red{\frac{1}{{3\ln(10)}}\frac{ \sigma_{ij}\sigma^{ij}}{{H}^2_0}\bigg|_{d_{L}  = 1 [\rm{Mpc}]} 
\approx   0.10\, [{\rm{mag}}].}
\label{eq:renormM2}
%=    \frac{2}{{9\log10}}f^2\sigma^2_{R}\,,
\end{eqnarray}
where $\sigma_{ij} = \partial_{i}  v_{ j } -\delta^{ij}\partial_i v_{ j }/3$  is the rate of shear deformation tensor experienced by the matter field. 
In order to obtain equation \eqref{eq:renormM2} in terms of the shear tensor, we have analytically continued the redshift dependence of the terms in equation \eqref{eq:renormM} to zero. This helps to appreciate the physical phenomenon at play here. That is the spacetime around the observer is tidally deformed due to the presence of inhomogeneous structures. Evaluating the ensemble average of the terms in equation \eqref{eq:renormM2} with a smoothing scale set at 1[Mpc] gives a result consistent with figure \ref{fig:differenceapparenet}. This is also consistent with the result presented in \cite{Umeh:2022prn} where a non-perturbative approach was used.

\subsection{Fitting the Alcock-Paczyński parameters}\label{sec:alcock-pay}

Similar to the CMB, the  plasma physics of acoustic density waves before decoupling imprints a characteristic length scale in the matter distribution. Alcock and Paczyński \cite{Alcock:1979mp} introduced a clever way to recover the true separation between galaxies which then allows to use the Baryon Acoustic Oscillation (BAO) as a standard ruler.  The observed separation could be decomposed into the radial  ${\alpha}_{\p} $  and the orthogonal ${\alpha}_{\perp } $  components~\cite{Anderson:2013oza}.  These are the observables that are measured by the large scale structures \cite{Nadathur:2019mct}.  ${\alpha}_{\p} $  and ${\alpha}_{\perp } $ are proportional to the Hubble rate and the area distance at the mean  redshift of the observation respectively. When ${\alpha}_{\p}$ and  ${\alpha}_{\perp }$ are  interpreted in terms of the background FLRW model 
\begin{eqnarray}\label{eq:backgroundalccok}
\red{\bar{\alpha}_{\p}(z_s)  = \frac{ {H}^{\rm{fid}}}{{H}} \,, \qquad {\rm{and }}\qquad 
\bar{\alpha}_{\perp }(z_s)   = \frac{\bar{d}_{A}}{\bar{d}^{\rm{fid}}_{A}}}
\end{eqnarray}
and evolved to today, it leads to a discrepant value for the Hubble rate. Here ${H}^{\rm{fid}}$ and $d^{\rm{fid}}_{A}$ are the fiducial Hubble rate and area distance.   

\begin{figure}[h]
\centering 
\includegraphics[width=100mm,height=70mm] {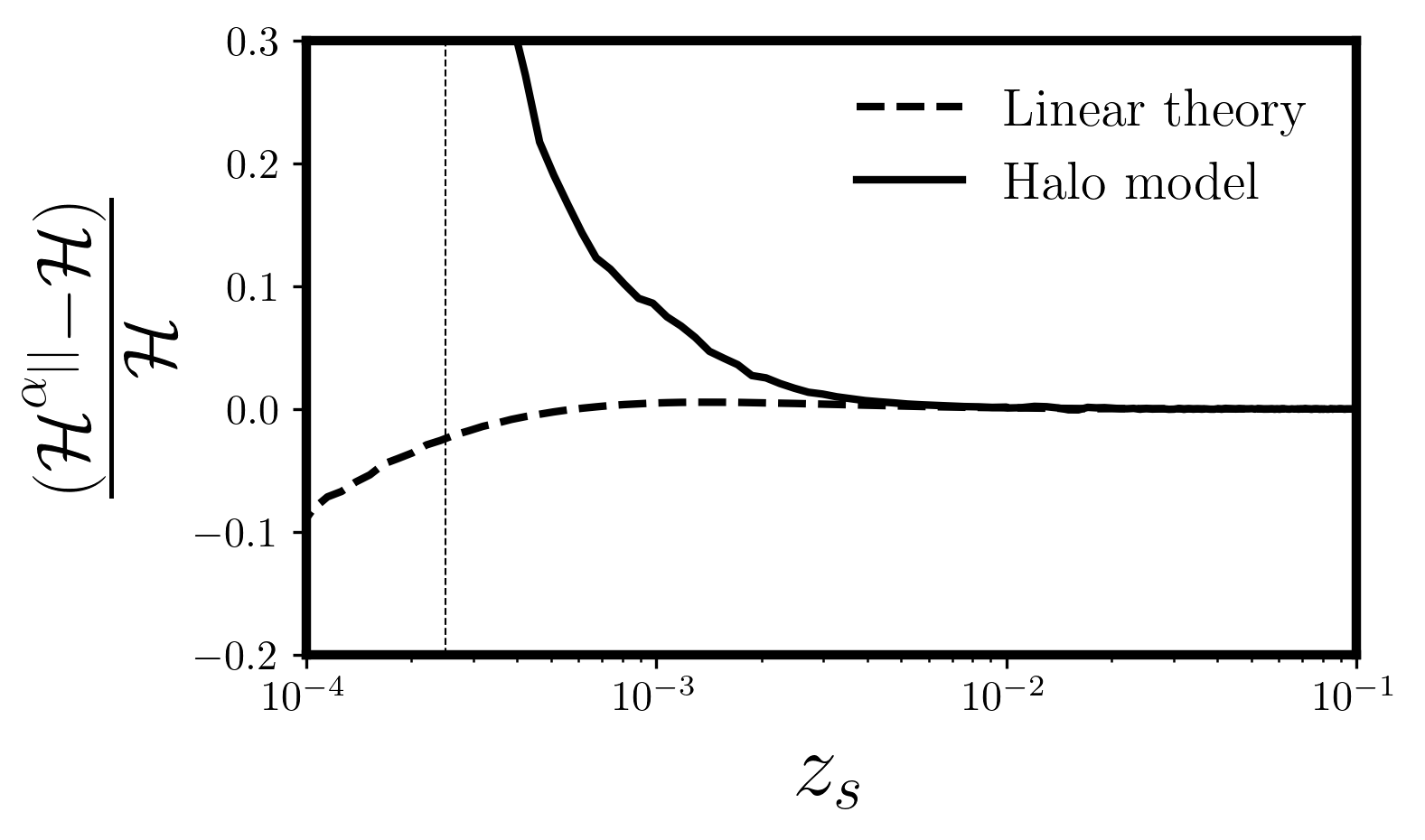}
\caption{\label{fig:fracdiffH}We show the fractional difference between the Hubble rate inferred from the radial component of BAO and the background FLRW spacetime prediction.\red{The vertical dashed line indicates the position of $z_{\rm{cut}}$. The dashed and thick lines indicate the cosmological perturbation theory and halo model predication for the matter power spectrum respectively. }
  }
\end{figure}

In a perturbed Universe, the radial  and orthogonal distances can be written as: $r{\p} = \bar{r}{\p} + \delta r_{\p}  = \bar{r}{\p} + \delta z/H$ and ${r}^A_{\perp} = \bar{r}^A_{\perp} + \delta r^A_{\perp}$,  where $\delta r_{\p}$ and $ \delta r^A_{\perp}$ are radial and tangential perturbations due to inhomogeneities respectively. Using these, the monopole of the generalised Alcock-Paczyński parameters in the presence of inhomogeneities  become~\cite{Umeh:2022hab}
\begin{align}
\averageA{\alpha_{ \p}(z_s)}  &
\simeq \frac{ {H} ^{\rm{fid} }}{{H}} \bigg[1+ \averageA{\frac{1}{H}\frac{\partial \delta z}{\partial r}}+ \mathcal{O}(\delta z)\bigg]\,,\label{eq:alphaparallel}
\\
\averageA{{\alpha}_{\perp } (z_s)} &=\frac{\averageA{{d}_{A}}}{d^{\rm{fid}}_{A}}\,.
\label{eq:alphabot}
\end{align}
Comparing equations \eqref{eq:alphaparallel} and \eqref{eq:alphabot} to their corresponding background FLRW limit (equation \eqref{eq:backgroundalccok}) and matching the corresponding terms  ($\averageA{\alpha_{ \p} } = \bar{\alpha}_{\p}$  and $ \averageA{\alpha_{\perp}} = \bar{\alpha}_{\perp })$, we find that the inferred Hubble rate from $\alpha_{\p}$
\begin{align}
\label{eq:eff-Hubble}
\HH^{\alpha\p}(z_s) &\approx \HH \left[1 -  \frac{1}{\HH^2_{s}}\left[\averageA{ \partial^2_{\|}{v_s} \partial^2\p{v_s}}+\averageA{\partial_{\|}{v_s}\partial^3\p{v_s}}- \averageA{ \partial_{\|}v_o\partial^3\p{v_s}} \right]\right]\,.
\end{align}
Expanding the correction in spherical harmonics, the leading order approximation at low redshift is given by
\begin{align}
\frac{\HH^{\alpha\p}(z_s) } {\HH} -1&\approx -
%\\ \nonumber &&\times
\sum_{\ell = 2}^{\ell_{\rm{max}}} (2\ell+1)\int \frac{ {\d} k}{2\pi^2}  k^2 P_{m}(k)(f(z_s)D(z_s))^2
\\ \nonumber  & \times
\bigg[  j''_{\ell}(kr_s) j''_{\ell}(kr_s)
+\bigg( j'_{\ell}(kr_s)  
- \frac{\HH_o}{\HH_s}\frac{D_of_o}{f(z_s) D(z_s)} j'_{\ell}(kr_o) \bigg)j'''_{\ell}(kr_s)\bigg]\,.
\end{align}
We show a plot of the fractional difference between the Hubble rate that would be inferred from the analysis of  radial component  of the BAO  and the global Hubble rate as function of the cosmological  redshift  in figure \ref{fig:fracdiffH}. The difference between the effective Hubble rate and the global Hubble rate begin to emerge at about the redshift of $z \le 0.01$. 
When the expression in equation \eqref{eq:eff-Hubble} is analytical continued to $z=0$, i.e., zero redshift limit both $H^{\alpha\p}$  and ${H_0^{{d}_A}}$ give the same result (see \cite{Umeh:2022hab,Umeh:2022prn} for further details).
 \begin{eqnarray}\label{eq:fracdiff}
\frac{({H_0^{{d}_A}}- {H}_{0})}{{H}_{0}}
 =- \lim_{z\to 0} \averageA{ \frac{\xi{\p}}{\bar{d}_{A}}} \approx  \frac{2}{15}\averageA{\frac{ \sigma_{ij}\sigma^{ij}}{ H^2_0}} = \frac{H^{\alpha\p}_0 -  H_0} {H_0}\,.
 \end{eqnarray}
Evaluating the ensemble average of equation \eqref{eq:fracdiff} at 1[Mpc] gives about 10\% difference.  This corresponds to the amount required to solve the Hubble tension.  Equation \eqref{eq:fracdiff} shows that the neighbourhood of the observer is curved and photons travel longer distance to get the observer. 
 %The inferred $H_0$ from both ${\alpha}_{\perp } $ is  given by

 \section{Breakdown of the expanding coordinate system}\label{sec:breakdown}

On the background FLRW spacetime one can easily extrapolate the magnitude-redshift relation down to zero redshift. In the presence of virialised structures (our local group), this is no longer possible~\cite{Umeh:2022prn}. This immediately constrains the map between the cosmological redshift and the monopole of the area distance shown in figure \ref{fig:BiasEv}. \red{It is possible to estimate the  length scale where the transition occurs given a halo model}.  \red{There are also pieces of evidence that the transition length scale have been detected  in the analysis of the large scale peculiar velocity survey, in this case, it is called the radius of the zero-velocity surface~\cite{1999AandARv...9..273V,Karachentsev:2008st}}. 

Given a geodesic deviation equation for time-like geodesics with the initial conditions at $\tau_{\rm{ini}} = 0$, zero-velocity surface is a conjugate point or caustic~\cite{Ellis:1998ha,Umeh:2022prn}. The appearance of a conjugate point indicates a breakdown of the global expansion coordinate chart~\cite{Witten:2019qhl}. 
Within the halo model, we can estimate this scale by calculating the comoving radius where the trace of the covariant derivative of the observer 4-velocity vanishes: $R_{0}= - ({c}/{3 H_0})  ({\d \ln \rho}/{\d \ln r})$, where $c$ is the speed of light and $\rho$ is the mass-density. 
 It is straightforward to calculate ${\d \ln \rho}/{\d \ln r}$ using any of the halo profiles~\cite{Diemer:2017bwl}. There are observational constraints for our local group, it is given by $R_0\sim(0.95- 1.05) [{\rm{Mpc}} ]$~\cite{1999AandARv...9..273V,Li:2007eg,Karachentsev:2008st}. In terms of the cosmological redshift, it corresponds to about $z_{\rm{cut}} = 2.4\times 10^{-4}$.  
 
 \red{The existence of $z_{\rm{cut}} $ implies that when the distance between two points in the universe is less than the radius of the zero-velocity surface ($|{\x}_2-{\x}_1| <R_{0}$), the global FLRW model  or the cosmological perturbation theory on an FLRW background cannot give the best fit model. }   The discussion on how to calculate the distance to the Milky-way cepheids or the Large Magellanic cloud and the impact of the heliocentric proper motion on the area distance would be presented elsewhere.

\section{Conclusions}\label{sec:conc}

\blue{This paper  shows} that the area distance at very low redshift limit includes a non-negligible correction due to nonlinear structure formation to the background FLRW spacetime predication. \blue{The } amplitude of the correction is about 10\% at $z =z_{\rm{cut}} = 2.4\times 10^{-4}$. The nonlinear correction is due to the effect of radial lensing or the Doppler lensing, \blue{i.e. a small scale lensing effects by galaxies orbiting about the centre of the  massive clusters.}  The leading order part of this effect  is  due to a correlation between observer peculiar velocity and redshift space distortion term at the source. \blue{This effect manifests as} a displacement in the apparent position of a source with respect to the peculiar velocity of the observer.  
%We referred to this as parallax effect.
 The implications of this effects on the apparent magnitude of the \blue{Sn~Ia} was studied in detail \blue{in sub-section \ref{sec:abstension}}.  The result show that
\blue{the Doppler lensing effects impact} the calibration of the absolute magnitude of the \blue{Sn~Ia} in the Hubble flow and that it could explain the supernova absolute magnitude tension without the need for any exotic early or late dark energy component.
%We discuss the contribution of the radial lensing effects to the variance in the Hubble residuals. 

Although the impact of the heliocentric observer relative velocity was neglected \blue{because  the cosmological perturbation theory used is not valid in the multi-stream region}.  This is the major reason the analysis reported here focused on the coordinate independent contribution of the effects of small scale inhomogeneities on the apparent magnitude of the \blue{Sn~Ia}. 
%We believe that  the heliocentric observer relative velocity will impact  the intercept of the Hubble diagram only. 
\blue{However, it is important to note that the impact or correction to} the supernova absolute magnitude will remain unchanged  when the impact of the heliocentric observer relative velocity is included.  \blue{In cosmological analysis, the impact of the heliocentric observer relative velocity could be calculated and subtracted off from the observed signal.}
%Our analysis is focused only on the coordinate independent part of the distance modulus.   

\blue{Using the expression for the area distance} that includes the radial/Doppler lensing effects, \blue{ the generalised Alcock-Paczyński parameters could be derived, in the paper it was shown that it is possible to} reconcile the inferred Hubble rate from  the BAO measurements  with  the Hubble rate from the analysis of the apparent magnitude of the \blue{SN~Ia} by  the SH0ES collaboration without any need for exotic late/early dark energy model. 
\blue{Including the effect of small scale non-linearity does not lead to any} difference between the TRGB and the SH0ES calibration of the absolute magnitude of the  \blue{Sn~Ia} in the Hubble flow.

\red{ Finally, there are several other proposed solutions to Hubble or  \blue{Sn~Ia} absolute magnitude tension.  These solutions usually  invoke exotic early or late dark energy model~\cite{Dainotti:2021pqg,Niedermann:2021vgd} (see \cite{Knox:2019rjx,Abdalla:2022yfr} for a list of all possible models within this framework). There are also frame-dependent dark energy model approach~\cite{Adler:2019fnp} (this approach explains the SH0ES result but gives  a large value of the Hubble constant from the BAO analysis). There is also proposals that explains the Hubble tension as a manifestation of the quantum measurement uncertainties~\cite{Capozziello:2020nyq}, or the impact of the  evolving gravitational constant leading to a modification of gravity at late-time~\cite{Zumalacarregui:2020cjh,Dainotti:2022bzg,Benevento:2022cql}. All these approaches except the frame dependent dark energy(frame dependent dark energy model breaks 4D diffeomorphism invariance)  assume that the FLRW background spacetime is valid on all scales and at all times. One unique about our approach is that it explains the Hubble tension within the standard model by simply including the effect of small scale perturbations. 
It will be interesting to see whether there is a projection of the hyperconical universe that explains both the dark energy and the Hubble tension better than the $\Lambda$CDM universe~\cite{Monjo:2017dyp,Monjo:2018upl}.
}

\section*{Acknowledgement}

I benefited immensely from the works of G.F.R. Ellis on this topic. 
OU  is supported by the 
UK Science \& Technology Facilities Council (STFC) Consolidated Grants Grant ST/S000550/1 and the South African Square Kilometre Array Project.  For the purpose of open access, the author(s) has applied a Creative Commons Attribution (CC BY) licence to any Author Accepted Manuscript version arising.
Most of the tensor algebraic computations in this paper were done with the tensor algebra software xPand~\cite{Pitrou:2013hga} and xTensor~\cite{xAct}.

\section*{Data Availability}
%Data sharing is not applicable to this article as no new data were created or analysed in this study.
Supporting research data are available on reasonable request from the  author.

\vspace{2cm}

%\section*{References}

% \bibliographystyle{iopart-num} 
% \bibliographystyle{$HOME/Dropbox/UWC_papers/Effective_fnl/Biassecondorder/JHEP}
%\bibliography{$HOME/Dropbox/UWC_papers/q-dipole/draft/cosmoref.bib}

\providecommand{\newblock}{}

\end{document}